\documentclass[reprint,aps,prb,amsmath, amssymb, amsfonts,superscriptaddress]{revtex4-2}
\usepackage{graphicx}
\usepackage{xcolor}
\usepackage{cancel}
\usepackage{soul}
\usepackage{braket}
\usepackage{calc}
\usepackage{bbold}
\usepackage{bm}
\usepackage{hyperref}

\newcommand{\up}{\uparrow}
\newcommand{\down}{\downarrow}
\newcommand{\Up}{\Uparrow}
\newcommand{\Down}{\Downarrow}

\begin{document}

\title{Entanglement Generation and Stabilization by Coherent Collisions}

\author{A. Mert Bozkurt}
\affiliation{QuTech and Kavli Institute of Nanoscience, Delft University of Technology, P.O. Box 4056, 2600 GA Delft, The Netherlands}

\author{Rosa L\'{o}pez}
\affiliation{Institute for Cross-Disciplinary Physics and Complex Systems IFISC (UIB-CSIC), E-07122 Palma de Mallorca, Spain}

\author{Sungguen Ryu}
\email{sungguen@ifisc.uib-csic.es}
\affiliation{Institute for Cross-Disciplinary Physics and Complex Systems IFISC (UIB-CSIC), E-07122 Palma de Mallorca, Spain}

\date{\today}

\begin{abstract}
 Collision is a useful tool for revealing quantum effects and realizing quantum informational tasks. 
 We demonstrate that repeated collisions by itinerant electrons can dissipatively drive two remote spin qubits into an entangled state in a generic collisional framework.
 A coherent spin exchange with either qubit facilitates entanglement generation.
 When combined with proper local driving, these collisions induce an entangled steady state in most collision configurations. 
 Particularly, the collision which is symmetric for the two qubits results in a unique steady state close to a maximally entangled state.
 Due to the dissipative nature of the process, the entanglement persists in the presence of decoherence, provided the collision frequency exceeds the decoherence rate.
 Our model can be experimentally implemented using single-electron sources. 
\end{abstract}

\maketitle

\section{Introduction}
Collisions are ubiquitous in physics and play a significant role in condensed matter systems, revealing quantum effects such as fermionic antibunching~\cite{liu_quantum_1998,olkhovskaya_shot_2008, bocquillon_coherence_2013}, 
charge fractionalization~\cite{wahl_interactions_2014,freulon_hong-ou-mandel_2015}, 
anyonic braiding effect~\cite{bartolomei_fractional_2020,lee_non-abelian_2022}, 
Coulomb repulsion effect at mesoscopic beam splitter~\cite{ryu_partition_2022, brange_interacting_2023,fletcher_time-resolved_2023, wang_coulomb-mediated_2023, ubbelohde_two_2023},
and quantum tomography~\cite{jullien_quantum_2014,bisognin_quantum_2019}.
Furthermore, collisions among helical electrons and nuclear spin memories have been proposed as mechanisms for topological information engines~\cite{bozkurt_work_2018,bozkurt2023Topological, bozkurt2024Entropy}.

Collisional models~\cite{ciccarello_quantum_2022} have been a useful theoretical tool for describing the dynamics of generic open quantum systems. 
In these models, the macroscopic environment is represented by a large collection of small units that sequentially collide with the system.
A recent model describes the environment using itinerant wave packets~\cite{jacob_thermalization_2021,jacob_quantum_2022,tabanera_quantum_2022,jacob_two-point_2023,tabanera-bravo_thermalization_2023}. 
The impact of these collisions on the system relies on the competition between the energy uncertainty of the packet and the level spacing of the system.
When the energy uncertainty is smaller, the collisions tend to thermalize the system.
Conversely, in the opposite regime, collisions induce coherence in the system, driving it out of thermal equilibrium.
This regime provides an opportunity to exploit coherent collisions as a resource for quantum information processing, such as realization of quantum gates~\cite{jacob_quantum_2022} and dissipative generation of entanglement
~\cite{kraus_preparation_2008,verstraete_quantum_2009, muschik_dissipatively_2011, krauter_entanglement_2011, stannigel_driven-dissipative_2012, benito_dissipative_2016, ryu_quantum_2022, bello_entangling_2022}.

Such a coherent collisional regime is experimentally feasible, thanks to single-electron sources~\cite{pekola_single-electron_2013,feve_-demand_2007, moskalets_quantized_2008, keeling_minimal_2006, dubois_minimal-excitation_2013,  hermelin_electrons_2011, kaestner_non-adiabatic_2015, giblin_towards_2012, ryu_ultrafast_2016,fletcher_continuous-variable_2019,freise_trapping_2020}. 
Based on AC voltages, these sources generate single-electron wave packets on demand. 
High-energy sources~\cite{giblin_towards_2012, kaestner_non-adiabatic_2015,ryu_ultrafast_2016,fletcher_continuous-variable_2019,freise_trapping_2020} generate packets with energy uncertainty as large as $\sim 1$ meV, exceeding the typical level spacing of quantum-dot spin qubits~\cite{burkard_semiconductor_2023}. 
The application of the single-electron sources to quantum information processing has been considered for realizing flying qubits~\cite{bauerle_coherent_2018, piccione_reservoir-free_2023} by itinerant electron excitations. 
This application seems promising, given that spin coherence lengths of order of $\mu$ms have been confirmed for such qubits~\cite{flentje_coherent_2017}.
Using collisions with itinerant electrons as a ancillary resource for quantum informational tasks has been considered, with the help of qubit initializations or post-selections~\cite{ciccarello_electron_2007,ciccarello_extraction_2008,ciccarello_reducing_2009,ciccarello_teleportation_2010,ciccarello_physical_2010}.
However, using the collisions for deterministic and dissipative generation of entanglement has received limited attention~\cite{benito_dissipative_2016}. 
The dissipative entanglement generation has a remarkable merit of being stable in the presence of decoherence due the dissipative nature.

In this paper, we show that itinerant electrons can mediate two remote spin qubits to generate and stabilize the entanglement in a generic collisional framework (see Fig.~1).
The collisions flip one of the qubits randomly but coherently, generating entanglement.
When combined with proper local drivings, the collisions induce an entangled steady state across most of the parameter space.
Weak, symmetric collisions for the two qubits result in a steady state with nearly maximal entanglement.
The entangled steady state remains stable even in the presence of the thermal relaxations and dephasing, as long as the collision occur at a frequency much higher than the decoherence rates.

The paper is organized as follows.
In Sec.~\ref{sec:model}, we present our model.  
In Sec.~\ref{sec:col}, we describe the dynamic mapping of the qubits which are collided with an electron using multiparticle scattering theory, which considers the spin exchange during the collision. 
In Sec.~\ref{sec:col2}, the dynamics when the collision and the local drivings are combined is described. 
In Sec.~\ref{sec:max_ent}, we analyze the steady state and find the condition to achieve the maximal entanglement.
In Sec.~\ref{sec:ent_gen}, the steady state entanglement for general collsional configurations are discussed.  
Finally, we discuss our results and conclude in Sec.\ref{sec:dis}.

\begin{figure}[t]
  \centering
  \includegraphics[width=\columnwidth]{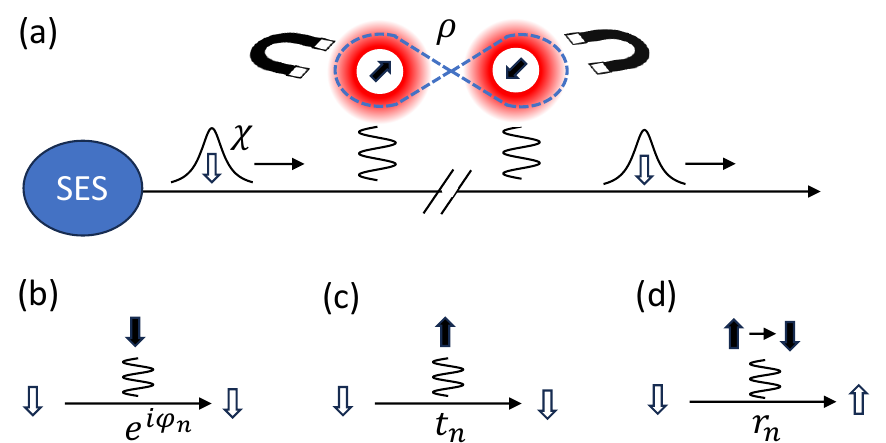}
  \caption{Collisional model for entanglement. (a) Distant two spin qubits, described by joint density matrix $\rho$, are collided repeatedly with electron excitations of spin state $\chi$ which are emitted into a chiral channel by a single-electron source (SES). 
  Each qubit is subject to a tunable magnetic field and a thermal bath.
  (b)--(d) Two-particle scattering matrix for a collision between an electron spin and $n$th ($n=1,2$) qubit when the electron spin is initially $\ket{\Down}$. 
  Electron [qubit] spin is denoted by the empty [filled] arrows. }
  \label{fig:setup}
\end{figure}

\section{Model}
\label{sec:model}
Our system consists of two distant spin qubits, each subject to local magnetic fields with tunable strength and orientations, as well as thermal baths (see Fig.~\ref{fig:setup}(a)). 
In addition, a single-electron source generates electron excitations that are injected into a one dimensional chiral channel at a frequency $f_{\text{col}}$.
This channel passes near the system, causing collisions between the electrons and the qubits, resulting in spin exchange.

We make the following assumptions.
The kinetic energy uncertainty of the initial electron is much larger than the level spacing of the qubits, preventing decoherence of the qubits by the collision~\cite{jacob_thermalization_2021}. 
The channel is assumed to be chiral, eliminating back-scattering, which can be achieved using quantum Hall edges.
We also assume a linear dispersion relation in the channel to avoid wave packet spreading, a reasonable approximation in typical single-electron source setups~\cite{emary_phonon_2016}.
Finally, we assume that the coherence length for the electron spin is larger than the distance between the qubits, allowing us to study the effect of coherent collisions.

The main ingredient to describe the collision is the two-particle scattering amplitudes~\cite{goorden_two-particle_2007} between an electron and a qubit $n$ ($=$1,2). 
We first consider the case that the electron spin is initially aligned to $-\hat{z}$, denoted by $\ket{\Down}$.
When the qubit is initially in $\ket{\down}$, no spin exchange occurs, and the electron scatters out, obtaining a forward-scattering phase $\varphi_n$, as described in Fig.~\ref{fig:setup}(b).
When the qubit is initially in $\ket{\up}$, 
the electron spin either remains unchanged with probability amplitude $t_n$, or flips with probability amplitude $r_n$ as shown in panels (c) and (d) of Fig.~\ref{fig:setup}, respectively.
The scattering amplitude for the initial electron spin $\ket{\Up}$ is determined by the schematics (b)--(d) with all the spins flipped, due to the spin-flip symmetry of the spin-spin interaction.
Hence, the two-particle scattering operator is written as
\begin{equation}
\begin{aligned}
s_n 
= &e^{i\varphi_n} \ket{\Up\up}\bra{\Up\up} + e^{i\varphi_n} \ket{\Down\down}\bra{\Down\down} \\
&+t_n \ket{\Up\down}\bra{\Up\down}  
 +r_n \ket{\Down\up}\bra{\Up\down} \\
&+t_n \ket{\Down\up}\bra{\Down\up}  
 +r_n \ket{\Up\down}\bra{\Down\up} .
 \end{aligned} 
 \label{eq:sn}
\end{equation}

As described in Appendix~\ref{A:Heisenberg}, 
the values of the amplitudes $t_n$,$r_n$ and phase $\varphi_n$ are determined by the microscopic details such as the strength of the electron-qubit interaction and the electron velocity.
Irrelevantly to the microscopic detail, we treat $t_n$,$r_n$, and $\varphi_n$ as model parameters, only restricted by the unitarity of $s_n$, namely $s_n^\dagger s_n=s_n s_n^\dagger =\mathbb{1}$.
The unitarity is equivalent to 
\begin{align}
    & |t_n|^2+|r_n|^2 = 1 ,  \label{eq:TR}
\\ 
    & \text{arg} (r_n/t_n) = \mp \pi/2 .
    \label{eq:tr}
\end{align}
Thus, we choose the global phase of the scattering amplitudes as $t_n=\sqrt{T_n}$, $r_n=\mp i\sqrt{1-T_n}$ without loss of generality.
The sign is $-$ when the electron-qubit interaction is ferromagnetic and $+$ when antiferromagnetic, see Appendix~\ref{A:Heisenberg}.
Throughout this paper, the upper and lower sign of $\mp$ (or $\pm$) in the analytic results refer to the ferromagnetic and antiferromagnetic cases, respectively. 
Furthermore, the spin conservation during the collision, $s_n^\dagger (\bm{\tau}+\bm{\sigma}_n) s_n= \bm{\tau}+\bm{\sigma}_n$ where $\bm{\tau}$ and  $\bm{\sigma}_n$ are the Pauli operators of the electron and qubit $n$ respectively, is equivalent to 
\begin{equation}
    \varphi_n= \mp \text{atan}\sqrt{(1-T_n)/T_n} \,.
    \label{eq:varphi}
\end{equation}
Thus, the probabilities $T_1$ and $T_2$ are the only independent model parameters which completely determines the collisions.
We remark that $T_n \approx 0$ and $T_n \approx 1$ correspond to strong collision (resulting in perfect spin exchange) and weak collision (no spin exchange), respectively.

\section{Dynamic mapping}
\subsection{Collisions}
\label{sec:col}
We first focus on the effect of the collision, assuming the absence of external magnetic fields and thermal baths, following the procedure of Ref.~\cite{jacob_thermalization_2021}.
The reduced density matrix $\rho$ of the two qubits changes due to a collision as
\begin{equation}
    \rho' = \mathbb{S} [\rho]  .
    \label{eq:rho'-col}
\end{equation}
The superoperator $\mathbb{S}$ is determined by evolving the joint state of the electron and qubits from just before to just after a collision and then tracing out the electron.
Specifically, the matrix elements of the superoperator, $\mathbb{S}_{j'k'}^{jk} \equiv  \text{Tr} (\ket{k'}\bra{j'} \mathbb{S} [\,\ket{j}\bra{k} \,] )$, are given by~\cite{jacob_thermalization_2021}
\begin{equation}
    \mathbb{S}_{j'k'}^{jk}
  = \braket{j'|\text{Tr}_{e} \Big[ S  
  \big( \rho_e \otimes \ket{j}\bra{k} \big) S^\dagger \Big] | k'} ,
  \label{eq:SS1}
\end{equation}
where $\rho_e $ is the density matrix of the initial electron describing both spatial and spin degrees of freedom, and the indexes $j$,$k$,$j'$ and $k' \in [\up\up, \up\down, \down\up, \down\down]$ denote the two-qubit basis states.
$\text{Tr}_e$ denotes tracing out the electron degrees of freedom.

\begin{figure}[t]
    \centering
    \includegraphics[width=\columnwidth]{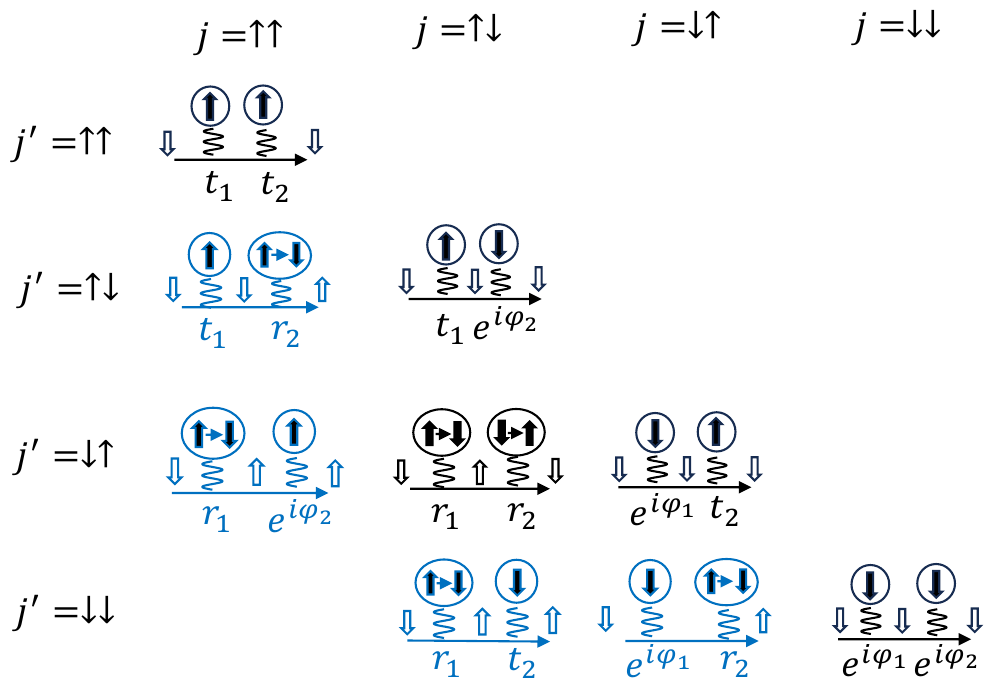}
    \caption{Three-particle scattering matrix, $\braket{\Down j'|S|\Down j} $ (black) and $\braket{\Up j'|S|\Down j} $ (blue). The matrix elements for the initial electron spin $\Up$ is determined through the spin-flip symmetry of $S$, namely
$\braket{\sigma' j'|S|\Up j} = \braket{\overline{\sigma'} \, \overline{j'}| S|\Down \overline{j}}$, where the overlines indicate the flip.  }
    \label{fig:S}
\end{figure}

$S$ is the three-particle scattering operator describing the scattering between one electron and two qubits.
Its matrix element $\braket{\sigma' j'|S|\sigma j} $, where $\sigma,\sigma' \in \{\Up, \Down\}$, describes the transition amplitude initially from $\ket{\sigma j}$ to  $\ket{\sigma' j'}$ after the scattering. 
Using that the three-particle scattering event is composed of two sequential two-particle scattering events, determined by Eq.~\eqref{eq:sn} and Fig.~\ref{fig:setup}(b)--(d)], and assuming interperiod independence, the three-particle scattering matrix is obtained as shown in Fig.~\ref{fig:S}.
The interperiod independence is assured because the electron emission period $f_{\text{col}}^{-1}$ (0.1--10 ns) is typically much larger than the time scale for the electron to complete the three-particle scattering (e.g., 10 ps for the electron velocity of $\sim 10^5$ m/s~\cite{kataoka_time--flight_2016} and qubit distance of 1$\mu$m).
Using the unitarity of the two-particle scattering, Eqs.~\eqref{eq:TR}--\eqref{eq:tr}, we confirm that the three-particle scattering matrix of Fig.~\ref{fig:S} also satisfies the unitarity, 
\begin{equation}
 S^\dagger S = S S^\dagger = \mathbb{1} .
 \label{eq:S-Unitarity}
\end{equation}
The spin conservation during the two-particle scattering, Eq.~\eqref{eq:varphi}, leads to the spin conservation during the three-particle scattering,
\begin{equation}
    S^\dagger (\bm{\tau}+\bm{\sigma}_1+\bm{\sigma}_2) S
    = \bm{\tau}+\bm{\sigma}_1+\bm{\sigma}_2 \,. 
    \label{eq:spin-cons}
\end{equation}

Using the condition that the energy uncertainty of the electron packet is much larger than the qubit level spacings, Eq.~\eqref{eq:SS1} is simplified to~\cite{jacob_thermalization_2021}
\begin{equation}
    \mathbb{S}^{jk}_{j'k'}
    = \sum_{\sigma,\sigma',\sigma''=\Up,\Down} 
    \hspace{-0.3cm} \braket{\sigma|\chi|\sigma'}
    \braket{\sigma'' j'|S|\sigma j} \braket{\sigma'' k'|S|\sigma' k}^* .
    \label{eq:SS2}
\end{equation}
Here, $\chi$ is the density matrix of the initial electron spin. 
In Eq.~\eqref{eq:SS2}, the terms with $\sigma\neq \sigma' $ describe the interference between two scattering events $\braket{\sigma'' j'|S|\sigma j}$ and $\braket{\sigma'' k'|S|\sigma' k}$.
The fact that the final electron spins of the two events are the same as $\sigma''$ reflects the decoherence in the qubits induced by being effectively measured via the electron spin.
Eqs.~\eqref{eq:rho'-col}, \eqref{eq:SS2}, and Fig.~\ref{fig:S} determine the evolution of the qubits through collisions with the electron. 

We first observe that entanglement is generated through collisions. 
For example, when the initial qubits in $\ket{\up \up}$ are collided with an electron spin $\ket{\Down}$,
either one of the two qubits is flipped. 
The specific qubit that flips is uncertain in a coherent manner, resulting in a state which is a superposition of $\ket{\up\down}$ and $\ket{\down\up}$ states, as illustrated in the first column of Fig.~\ref{fig:S}.
However, this entanglement is not stable as repeated collisions eventually drive the qubits to become polarized, aligning their spins with the electron spins.

\subsection{Collisions with local driving}
\label{sec:col2}
Now we consider the collisions in the presence of the external magnetic fields and the baths.
The three-particle scattering is completed in a very short time (typically tens of picoseconds) as discussed above.
Hence, during scattering, the effects of Larmor precession and thermal fluctuations are negligible.
This leads to the change in the qubits during one period, $1/f_{\text{col}}$, being determined by the sequential acts of the scattering and the evolution in between the scatterings:
\begin{equation}
    \rho' = e^{\mathbb{L}/f_{\text{col}}} \big[\mathbb{S} [\rho] \big] .
    \label{eq:ME}
\end{equation}

Here, $\mathbb{L}$ is the Liouvillian superoperator describing the time evolution of the two qubits under the influence of magnetic fields and thermal fluctuations:
\begin{equation}
\begin{aligned}
    \mathbb{L}[\rho]
    &= -\frac{i}{\hbar} [H, \rho]    
    + \sum_{\substack{n=1,2\\ \lambda=e,a}} \Gamma_{n \lambda}
    \Big( L_{n\lambda} \, \rho \, L_{n\lambda}^{\dagger}
    -\frac{1}{2}\big\{ L_{n\lambda}^\dagger L_{n\lambda} , \rho \big\} \Big) \\
    & \qquad +\sum_{n=1,2} \gamma_{n} \Big( (\hat{B}_n  \cdot \bm{\sigma}_n) \, \rho \,  (\hat{B}_n  \cdot \bm{\sigma}_n) -\rho\Big) .
\end{aligned}
\label{eq:L}
\end{equation}
The first term describes the unitary evolution due to the Hamiltonian of the qubits,
\begin{equation}
    H = -  \sum_{n=1,2} \frac{\hbar\Omega_n}{2} \hat{B}_n \cdot  \bm{\sigma}_n,
    \label{eq:H}
\end{equation}
where $\hat{B}_n$ is the unit vector describing the direction of the magnetic field applied to the $n$th qubit and $\Omega_n$ is the corresponding Larmor angular frequency.
The second term of Eq.~\eqref{eq:L} describes the thermal fluctuations induced by the baths.
$L_{n e}$ is the jump operator describing the spontaneous emission of the $n$th qubit:
\begin{align}
    L_{1e} 
    &= \big( \ket{\hat{B}_1;+}\bra{\hat{B}_1; -} \big) \otimes \mathbb{1}  ,\\
    L_{2e} 
    &= \mathbb{1} \otimes \big( \ket{\hat{B}_2;+}\bra{\hat{B}_2; -} \big)   ,
\end{align}
where $\ket{\hat{B}_n;+}$ and $\ket{\hat{B}_n;-}$ are the spinors directed to $\hat{B}_n$ and $-\hat{B}_n$, respectively.
$L_{na}=L_{ne}^\dagger$ is the jump operator for the absorption process.
The rates of the spontaneous emission and absorption, $\Gamma_{ne}$ and $\Gamma_{na}$ respectively, are related by the local detailed balance~\cite{esposito_three_2010}
\begin{equation}
    \frac{\Gamma_{na}}{\Gamma_{ne}} = \exp \Big[-\frac{\hbar\Omega_n}{k_B \mathcal{T}} \Big] .
    \label{eq:LDB}
\end{equation}
Here, $\mathcal{T}$ is the temperature of the baths (assumed to be the same in the two baths for simplicity). 
The third term of Eq.~\eqref{eq:L} describes pure dephasing~\cite{schaller_open_2014} with rate $\gamma_n$ induced by the fluctuations of the magnetic field. 

Equations ~\eqref{eq:ME}--\eqref{eq:LDB} allow for the numerical calculation of the time evolution of the system spins.
The steady state can be directly obtained through finding the eigenstate of the superoperator $e^{\mathbb{L}/f_\text{col}} \mathbb{S}$ with eigenvalue of 1.
The steady state is unique when there is no degeneracy for the unit eigenvalue.

\section{steady state entanglement}
\label{sec:std_ent}
We now study the conditions required to obtain a steady state entanglement.
We find that a steady state close to a maximally entangled state can be formed when the collision strengths of each qubit are equal, i.e. $T_1=T_2$.
Since these strengths may not always be controllable, we also investigate the conditions needed to maximize steady state entanglement for general values of $T_1$ and $T_2$.

\subsection{Maximal entanglement }
\label{sec:max_ent}

The necessary conditions for achieving maximal steady state entanglement are as follows:
(i) Entanglement requires that the two qubits communicate beyond local operations and classical communications~\cite{mintert_measures_2005}. This implies that the initial electron spin must be coherent and the spin transfer from the location of one qubit to another should be coherent. 
(ii) Maximal entanglement requires that the steady state should be symmetric to the qubit exchange.  Hence, the collisional map $\mathbb{S}$ and the Liouvillian $\mathbb{L}$ and should also have the exchange symmetry. This leads to $T_1=T_2\equiv T \approx 1$ and $\vec{B}_1=\vec{B}_2 $.
Note that the weak collision condition, $T\approx 1$, is necessary due to the chiral geometry of the collisional setup. 
The effect of the collision on the second qubit differs from that on the first qubit because the incident electron spins in the two cases are different, unless the collisions are weak enough that the electron exchanges spin with only one of the two qubits.
(iii) The reduced density matrix of the first qubit should be the white-noise state $\mathbb{1}/2$. 
This requirement and the chirality of the channel imply that the magnetic field $\vec{B}_1$ should flip the qubit in between the collisions to prevent the qubit from being polarized along the initial electron spin. 
Indeed, the first qubit becomes the white-noise state, when $\vec{B}_1$ perpendicular to the initial electron spin and $\Omega_1 f_{\text{col}}^{-1} = \pi$, see Appendix~\ref{A:rho1}.

\begin{figure}[t]
  \centering
  \includegraphics[width=1\columnwidth]{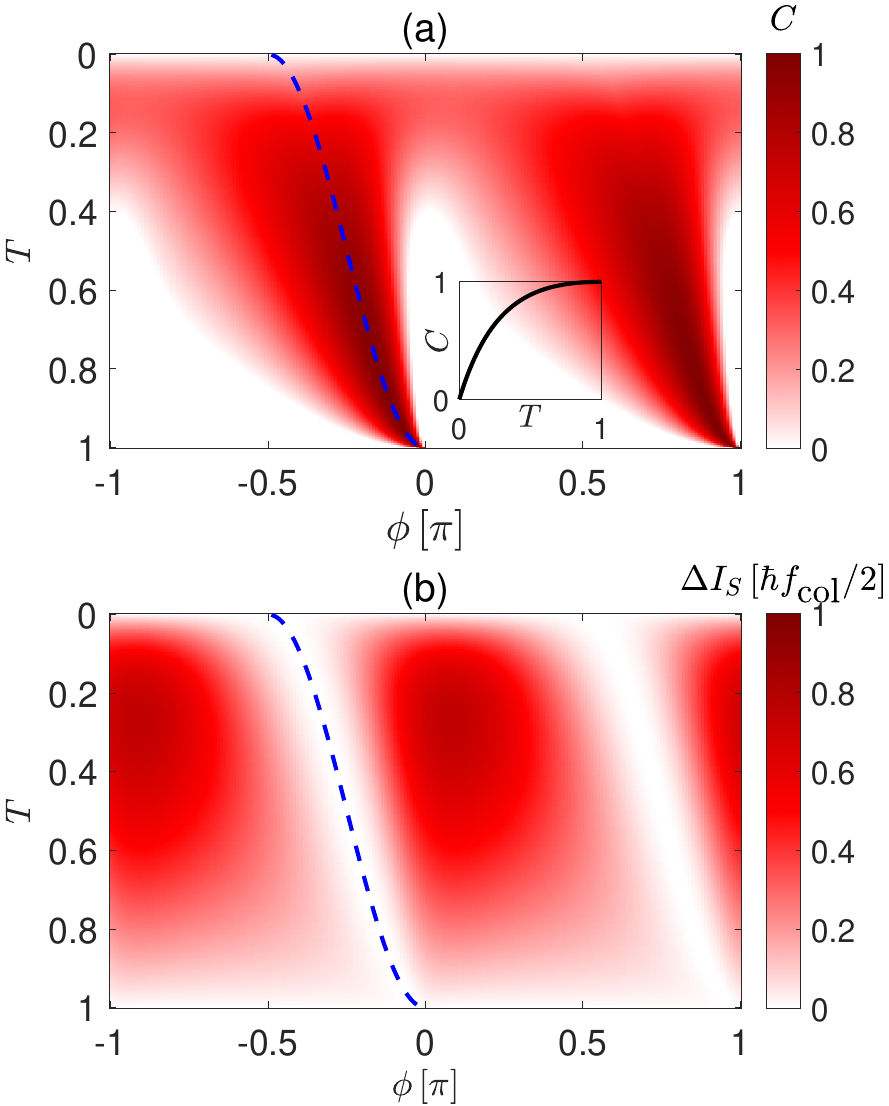} 
  \caption{Steady state with maximal entanglement.
  (a) Concurrence. 
   Blue dashed line indicates the optimal condition, $\phi=\varphi$.
   Inset: the concurrence at the optimal condition.
  (b) Spin current (along $\hat{z}$) of electrons in the output channel subtracted by that of input channel. 
  Parameters: $\chi = \ket{\Down}\bra{\Down}$, the magnetic fields are determined by Eqs.~\eqref{eq:B1}--\eqref{eq:Omg}, 
  the collision is symmetric, i.e., $T_1=T_2\equiv T$,  
  and the decoherence is absent, i.e.,  $\Gamma_{ne}=\Gamma_{na}=\gamma_{n}=0$. The electron-qubit interaction is assumed to be ferromagnetic, hence  $\varphi= - \text{atan}\sqrt{(1-T)/T}$.}
  \label{fig:ent_phi}
\end{figure}

Interestingly, we find that relaxing condition (ii) slightly by allowing an azimuthal angle $\phi$ between $\hat{B_1}$ and $\hat{B_2}$ is critical.
The magnetic fields are described by
\begin{align}
    \hat{B}_1 &= \hat{x} ,  
    \label{eq:B1}\\
    \hat{B}_2 &= \cos \phi \,\hat{x} +\sin \phi \,\hat{y} .\label{eq:B2} \\
    \Omega_1 &= \Omega_2 = \pi f_{\text{col}} .\label{eq:Omg}
\end{align}
Here we have chosen $\chi=\ket{\Down}\bra{\Down}$ (among any pure spinor) and $\hat{B}_1=\hat{x}$ (among any direction perpendicular to the electron spin), without loss of generality.
Fig.~\ref{fig:ent_phi}(a) shows the steady state entanglement, quantified by concurrence $C$~\cite{horodecki_quantum_2009}.
When the azimuthal angle equals the forward scattering phase, namely $\phi=\varphi$ where $\varphi\equiv \varphi_1=\varphi_2$, the steady state entanglement becomes nearly maximal for weak collisions, i.e., $T\rightarrow 1$. 
On the other hand, when $\phi=0$ the steady state entanglement is drastically weaker, being $0$ for $0<T<0.5$ and $\lesssim 0.1$ otherwise.

To understand the importance of the condition $\phi=\varphi$, it is convenient to express the dynamic mapping in a singular value decomposition. The dynamic map describing the collision and Larmor precession is determined by two operators $\hat{t} \equiv e^{-iH/(\hbar f_{\text{col}})} \braket{\Down|S|\Down}$ and $\hat{r} \equiv e^{-iH/(\hbar f_{\text{col}})} \braket{\Up|S|\Down}$ as
\begin{equation}
    e^{\mathbb{L}/f_{\text{col}}} [\mathbb{S}[\rho]] 
    = \hat{t} \rho \, \hat{t}^\dagger +\hat{r}\rho \, \hat{r}^\dagger .
\end{equation}
The operator $\hat{t}$ ($\hat{r}$) describes the evolution given that the electron spin has not flipped (flipped). 
The operators $\hat{t}$ and $\hat{r}$ are expressed in their singular decompositions as (see Appendix~\ref{A:SVD} for the derivation),
\begin{figure}[t]
    \centering
    \includegraphics[width=\columnwidth]{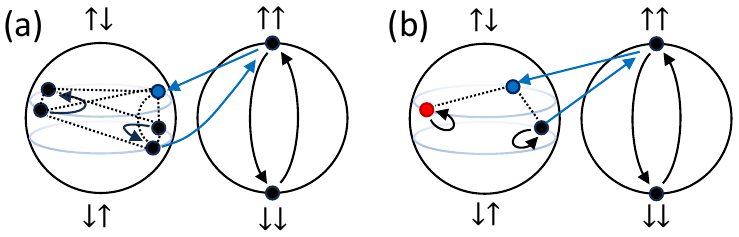}
    \caption{Schematics of the dynamic mapping for $\phi=0$ (a) and $\phi=\varphi$ (b). The initial and final states of the mapping are shown in Bloch spheres of basis of $\{\ket{\up\down}, \ket{\down\up}\}$ and $\{\ket{\up\up}, \ket{\down\down}\}$. 
    Black and blue arrows denote the transition by $\hat{t}$ and $\hat{r}$, respectively [see Eqs.~\eqref{eq:tt} and \eqref{eq:rt}].
    Dotted lines refer to the hopping through spinor overlap. 
    In the case of $\phi=0$, the steady state is formed by statistical ensemble of many pure states (denoted as black and blue points).
    In the optimal condition, $\phi=\varphi$, the steady state is formed by a unique pure entangled state (denoted as the red point having polar angle $\vartheta_T$ and azimuthal angle $\pi-\varphi$). 
    Here, the electron-qubit interaction is assumed to be weak and ferromagnetic, namely $\vartheta_T \approx \pi/2$ and $ \varphi \lesssim 0 $. $\vartheta_T$.}
    \label{fig:sch_DM}
\end{figure}
\begin{align}
 \hat{t}
 =&-e^{i(2\varphi-\phi)} \ket{\up\up} \bra{\down\down}
 - T e^{i\phi} \ket{\down\down}\bra{\up\up} \nonumber   \\
 & -T e^{i\phi}  \ket{\Psi(\pi -\vartheta_T, \varphi-2\phi)} \bra{\Psi(\pi -\vartheta_T, -\varphi)} \nonumber
 \\
 & +e^{i\phi} \ket{\Psi(\vartheta_T,\pi+\varphi-2\phi} \bra{\Psi(\vartheta_T, \pi-\varphi)} ,
\label{eq:tt} \\
 \hat{r}
 =& \pm i e^{-i\phi}\sqrt{1-T^2}  \ket{\up\up} 
 \bra{\Psi(\pi-\vartheta_T, -\varphi)} \nonumber \\
 & \pm i e^{i(\varphi+\phi)} \sqrt{1-T^2}  
 \ket{\Psi(\vartheta_T, -\varphi-2\phi)}\bra{\up\up} .
\label{eq:rt}
\end{align}
Here $\vartheta_T\equiv 2\, \text{atan} (\sqrt{T})$ is an angle describing the collision strength; $\vartheta_T $ changes from $\pi/2$ to $0$ as the collision strength is increased.
$
    \ket{\Psi(\vartheta, \varphi)  }  \equiv
    \cos(\vartheta/2) \ket{\up\down}
    +\sin(\vartheta/2) e^{i \varphi} \ket{\down \up}
$ is a state which has the polar angle $\vartheta$ and azimuthal angle $\varphi$ in the Bloch sphere of the subspace spanned by $\ket{\up\down}$ and $\ket{\down \up}$.

Fig.~\ref{fig:sch_DM} illustrates the schematics of the dynamic mapping by the collisions and the Larmor precession, as described by Eqs.~\eqref{eq:tt} and \eqref{eq:rt}. 
When $\phi=0$ (see Fig.~\ref{fig:sch_DM}(a)), the mapping leads a product state $\ket{\up\up}$ to an entangled state $\ket{\Psi(\vartheta_T, -\varphi)}$. However, this state is not stable and transferred to various states, such as $\ket{\Psi(\vartheta_T, \pi+\varphi)}$, $\ket{\Psi(\pi-\vartheta_T, \varphi) }$, $\ket{\up\up}$, and $\ket{\down\down}$, 
due to the mapping and hopping through spinor overlap. 
Consequently, the steady state is formed as an ensemble of these states, which does not exhibit entanglement as a whole.

In contrast, when $\phi=\varphi$ (see Fig.~\ref{fig:sch_DM}(b)), the structure of the transition paths is simplified, leading to a unique stable point at $\ket{\Psi(\vartheta_T, \pi -\varphi)}$.
Consequently, an arbitrary state eventually converges to this state, forming a pure steady state given by
\begin{equation}
\rho_{\text{st}}
 = \frac{\ket{\up\down} - e^{-i \varphi}\sqrt{T}\ket{\down\up} }{\sqrt{1+T}}   
(\text{h.c.}) , 
\label{eq:rhoSt-maxEnt}
\end{equation}
where (h.c.) denotes to the Hermitian conjugate of the preceding term. 
This state exhibits a concurrence value very close to 1 in the weak collisional limit, namely $T\approx 1$,
\begin{equation}
    C(\rho_{\text{st}})=\frac{4 T}{(1+T)^2} .
    \label{eq:C-opt}
\end{equation}
Eq.~\eqref{eq:C-opt} is consistent with the numerical results shown in the inset of Fig.~\ref{fig:ent_phi}(a).

\begin{figure}[t]
  \centering
  \includegraphics[width=0.9\columnwidth]{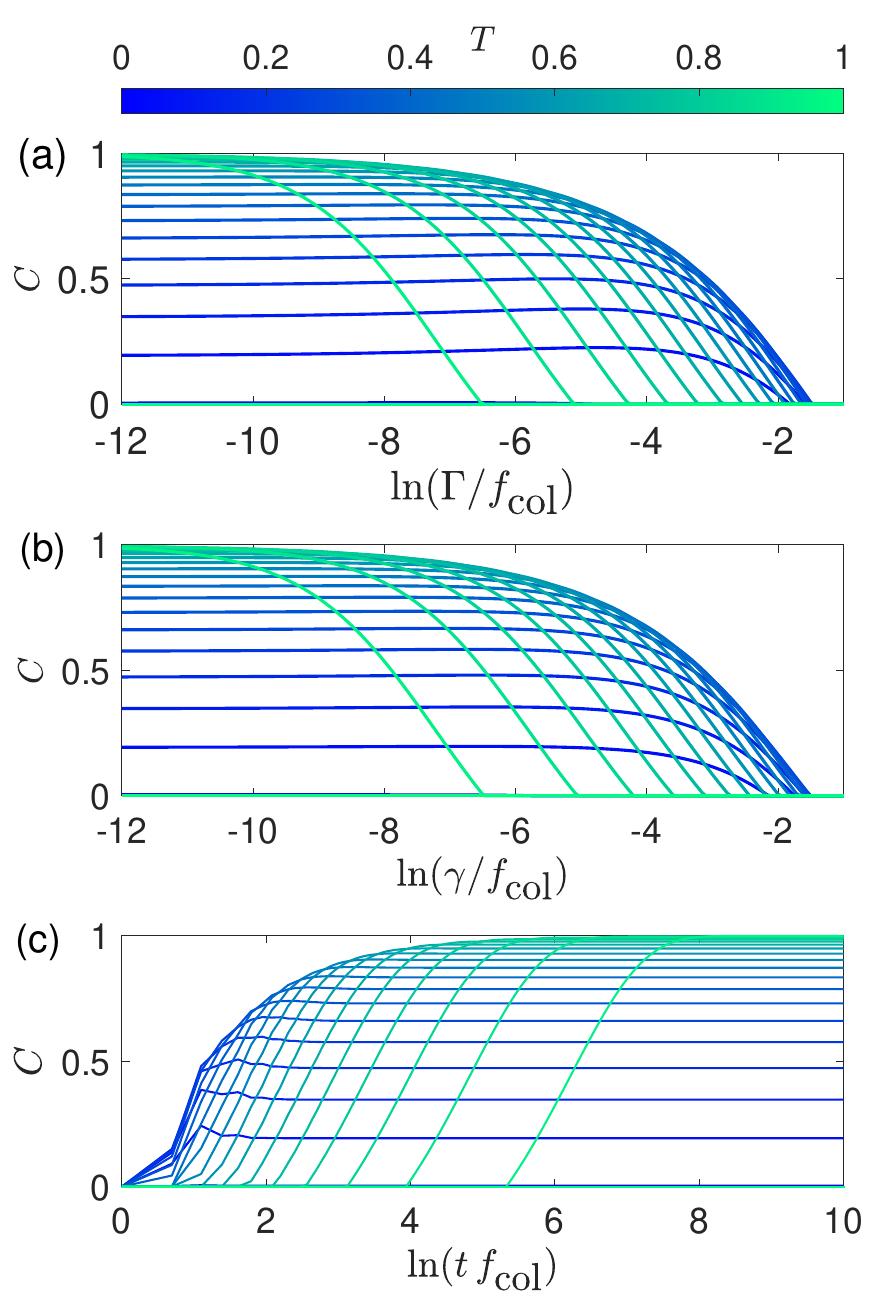} 
  \caption{Steady state entanglement with relaxation (a) and dephasing (b), and entanglement dynamics (c). The results are shown for various $T$ (color-coded).
  Parameters are the same as Fig.~\ref{fig:ent_phi} with the optimal condition, $\phi=\varphi$, except: In (a), $\Gamma_{1a}=\Gamma_{1e}=\Gamma_{2a}=\Gamma_{2e} \equiv \Gamma$ which corresponds to high-temperature limit $k_B \mathcal{T} \gg \hbar\Omega_1, \hbar\Omega_2$ and $\gamma=0$. In (b), $\gamma_1=\gamma_2 \equiv \gamma$ and $\Gamma=0$.
  In (c), $\Gamma=\gamma=0$ and the initial two-qubit state is chosen as $\rho = \mathbb{1}/4$. 
  }
  \label{fig:ent_time} 
\end{figure}

The forward-scattering phase $\varphi$ is not initially known in experiments, because it is determined by the type and strength of the collision interaction, as seen in Eq.~\eqref{eq:varphi}.
However, the optimal condition of the magnetic fields, $\phi=\varphi$, can be easily tuned experimentally without prior knowledge of the collision specifics.
As discussed above, the purity of the steady state, $\text{Tr} \, \rho_{\text{st}}^2$, reaches unity under the optimal condition and is smaller otherwise.
In analogy to the dark state in photonic setups~\cite{stannigel_driven-dissipative_2012}, the qubits in this pure steady state do not exchange spin with the electron.
Indeed, one can verify that the amplitude of transition $\braket{\Up|S|\Down}$ applied to Eq.~\eqref{eq:rhoSt-maxEnt} is zero. 
Therefore, $\phi$ can be tuned to its optimal value by measuring the electron spin current along $\hat{z}$ in the output channel and adjusting $\phi$ to minimize changes in the electron spin current between the input and output channels. 
Fig.~\ref{fig:ent_phi}(b) demonstrates this behavior, where the spin current is obtained using the spin conservation, Eq.~\eqref{eq:spin-cons}, and by calculating the spin expectation values of $\rho_{\text{st}}$, Eq.~\eqref{eq:rhoSt-maxEnt}, and $\mathbb{S}[\rho_{\text{st}}]$.

The steady state entanglement persists even in the presence of decoherence when its rate is much slower than the collision rate, as illustrated in
Figs.~\ref{fig:ent_time}(a) and (b).
The threshold of the decoherence rate, up to which the steady state remains entangled, decreases for weaker collisions. 
The behavior can be understood from the entanglement dynamics, see Fig.~\ref{fig:ent_time}(c).
As $T$ approaches 1 (indicating weaker collisions), the generation of entanglement slows down.   
As a result, decoherence more effectively destroys entanglement near the weak collisional limit.

Interestingly, the concurrence $C$ in Fig.~\ref{fig:ent_time}(c) for small $T$ shows an oscillation in early collisions.  We find that the oscillation is anticorrelated with an oscillation of probability of the qubits being in $\ket{\up \up}$. Namely, the concurrence reduction by a collision is accompanied with the increase of the probability. This is because that the transition from $\ket{\up\up}$ to a superposition of $\ket{\up\down}$ and $\ket{\down \up}$ (upper blue arrow in Fig.~\ref{fig:sch_DM}(c)) and the opposite transition (lower blue arrow) coexist in the early collisions. Such oscillation is less pronounced in the larger $T$ as the collision is weak and qubits evolve in time smoothly.

  To estimate the time scale on which the steady state is approached, we have numerically searched the time at which the concurrence $C$ reaches the half of the converged value in Fig.~\ref{fig:ent_time}(c). We find that the time scale is well described by $(1-T)^{-2} f_{\text{col}}^{-1}. $ Hence, the Bell state generation suggested by Eqs.~\eqref{eq:rhoSt-maxEnt} and \eqref{eq:C-opt} in the limit of $T \rightarrow 1$ should be understood as asymptotic generation with diverging time scale. As $T$ approaches 1, the interaction between the electron and qubits becomes weak, and hence the time scale on which the steady-state is reached grows divergently as $(1-T)^{-2} f_{\text{col}}^{-1}$.  Note that nearly maximal entanglement can be achieved in accessible time scale, e.g., $C=0.9972$ in $(1-T)^{-2} f_{\text{col}}^{-1}\sim 100$ ns time scale for $f_{\text{col}}=1 $GHz~\cite{giblin_towards_2012} and $T=$0.9.

\begin{figure}[t]
  \centering
  \includegraphics[width=0.8\columnwidth]{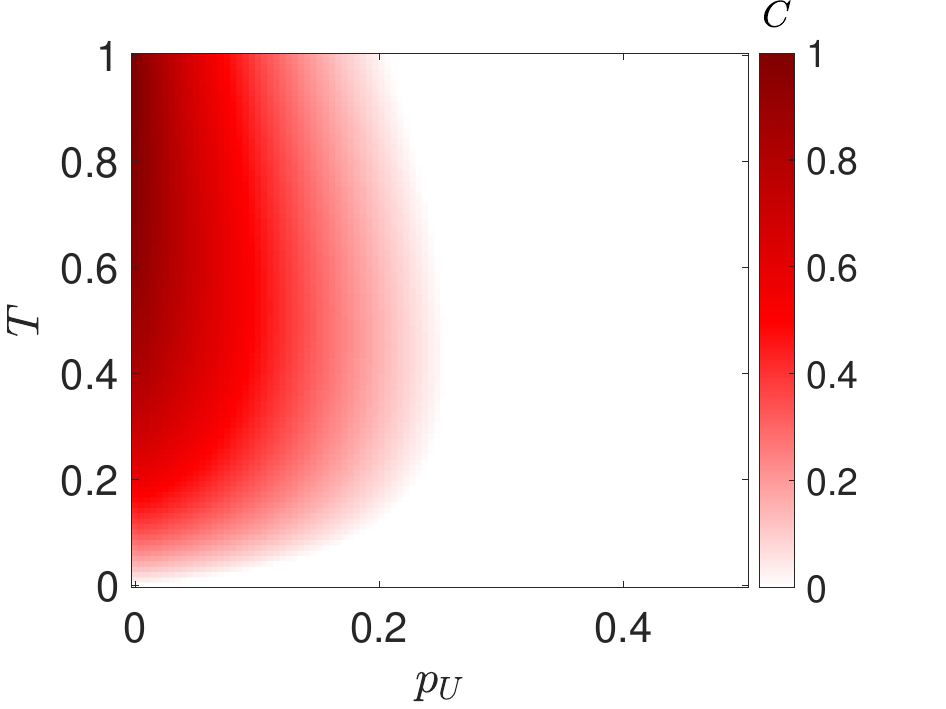} 
  \caption{Steady state entanglement with various electron spin coherence.  Parameters are the same as in Fig.~\ref{fig:ent_phi} with the optimal condition $\phi=\varphi$ except that $\chi = p_U\ket{\Up}\bra{\Up} +(1-p_U) \ket{\Down}\bra{\Down}$.  }
  \label{fig:ent_pU}
\end{figure}

The coherence of electron spins is crucial for the steady state entanglement.
Fig.~\ref{fig:ent_pU} demonstrates the concurrence when the initial electron spin is an incoherent mixture of up and down spins, described by
$\chi = (1-p_U) \ket{\Down}\bra{\Down} +p_U\ket{\Up}\bra{\Up}$.
As the probability $p_U$ to occupy the up-spin changes from 0 (corresponds to the fully coherent case) to 0.5 (fully incoherent case), the steady state entanglement decreases and eventually vanishes.
Furthermore, the coherence of electron spin between the collisions with each qubit is also crucial. There are two mechanisms which determine the spin coherence length of the itinerant electron. One is dephasing due to hyperfine interaction provided by nuclear spins. The other is spin-flip due to spin-orbit coupling. Due to the fast speed of the itinerant electron (which range from 4000 m/s \cite{hermelin_electrons_2011} to $10^5$ m/s \cite{kataoka_time--flight_2016} depending on the type of single-electron sources), the coherence lengths of both mechanisms are expected to be larger than micrometers~\cite{flentje_coherent_2017,stotz_coherent_2005,bertrand_fast_2016}.   
If the spin coherence length is shorter than the distance between the qubits, the collision changes $\ket{\up\up}$  to a classical ensemble of $\ket{\up\down}$ and $\ket{\down\up}$, rather than their superposition.
Hence the concurrence is expected to decrease with increasing qubit distance per coherence length. Furthermore, if the itinerant electron has experienced spin rotations due to the spin-orbit interactions, the optimal condition of the local magnetic fields will be different than the one found in our work.    

In the result of Fig.~\ref{fig:ent_phi}, we assume a ferromagnetic interaction between the electron and qubit. 
The results, including the concurrence and the electron spin current, of antiferromagnetic interaction are the same as the ferromagnetic results to which $\phi \rightarrow -\phi$ is applied.
The optimal condition $\phi=\varphi$ is independent of the interaction type.

\subsection{General case}
\label{sec:ent_gen}

In experimental settings, controlling the strength of electron collisions with each qubit might not be straightforward. On the other hand, local unitary operations on qubits are typically well-controlled. Therefore, it is desirable to study how much steady state entanglement can be achieved by tuning these local unitary operations under given collision strengths.

\begin{figure}[t]
  \centering
  \includegraphics[width=0.8\columnwidth]{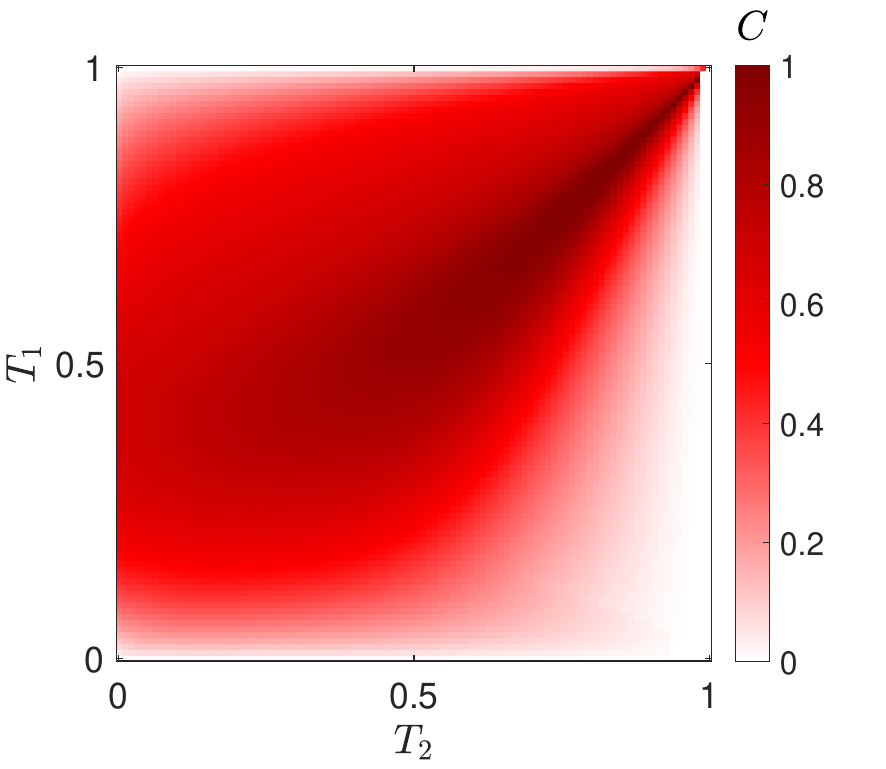} 
  \caption{Steady state entanglement for general $T_1$ and $T_2$. At each point, the magnetic fields are optimized to maximize the entanglement. The result of the optimized magnetic fields are given in Fig.~\ref{fig:pam_ent_T1T2}.
  Here we assume the absence of relaxation and dephasing, ferromagnetic electron-qubit interaction, and $\chi=\ket{\Down}\bra{\Down}$.
  }
  \label{fig:ent_T1T2}
\end{figure}

Figure~\ref{fig:ent_T1T2} shows the maximum steady state entanglement achieved by numerically optimizing the directions of the local magnetic fields, $\hat{B}_{i=1,2}= \sin(\theta_i) \cos(\phi_i)\hat{x} +\sin(\theta_i)\sin(\phi_i)\hat{y}+ \cos(\theta_i) \hat{z}$, and the strengths $\Omega_i$.
These fields induce corresponding Larmor precessions during the collision period $1/f_{\text{col}}$, enabling arbitrary local unitary operations.
Assuming $\chi = \ket{\Down}\bra{\Down}$, the entanglement depends on the difference in azimuthal angles, $\phi_2-\phi_1$, rather than their individual values.
The optimal parameters are given in Appendix Fig.~\ref{fig:pam_ent_T1T2}.
 The absence of spin relaxation and dephasing is assumed for the itinerant electrons.

Remarkably, significant steady state entanglement is observed across a wide range of $T_1$ and $T_2$.
This implies that in experiments where $T_1$ and $T_2$ can be measured, one can nearly always adjust local unitary operations to achieve an entangled steady state, guided by the optimal parameters given by Fig.~\ref{fig:pam_ent_T1T2}.
The strengths of collisions $T_1$ and $T_2$ can be extracted in experiments by measuring the electron spin current.
When strong magnetic fields are applied to the qubits along the opposite direction of the initial electron spin polarization, specifically when $\hat{B}_1=\hat{B}_2=\hat{z}$, $\chi=\ket{\Down}\bra{\Down}$, and $\Omega_1,\Omega_2 \gg k_B \mathcal{T}/\hbar$, the electron spin currents (measured along $\hat{z}$) at the channel between the qubits, $I_S'$, and at the output channel, $I_S''$, are given by
\begin{align}
    I_S'
    &= \frac{\hbar f_{\text{col}}}{2} (1-2 T_1) , \label{eq:Is'} \\
    I_S''
    &= \frac{\hbar f_{\text{col}}}{2} (1-2 T_1 T_2) , 
    \label{eq:Is''}
\end{align}
respectively. Thus, by measuring the spin currents $I_S'$, $I_S''$ and applying Eqs.~\eqref{eq:Is'} and \eqref{eq:Is''}, one can determine $T_1$ and $T_2$.  Due to the chirality of the channel, Eq.~(\ref{eq:Is'}) is valid irrelevantly to what happens in the downstream. Therefore, to measure $I_S'$ conviniently, one can divert the channel using a potential barrier or quantum point contact, and measure the spin current adopting a method convenient in experimental platform~\cite{burkard_semiconductor_2023}. 

Further analysis of the entanglement and optimal conditions reveals that when  $T_1 \approx T_2$, the optimal magnetic fields remain almost the same as the ones found in Sec.~\ref{sec:max_ent}: specifically $\Omega_1=\Omega_2 \approx \pi f_{\text{col}}$, $\theta_1=\theta_2 \approx \pi/2 $, and $\phi_2-\phi_1=(\varphi_1+\varphi_2)/2$.
Hence, if $T_1\approx T_2$ can be assumed without precise knowledge about their specific values, one can tune $\phi$ by minimizing the change of the electron spin current, as discussed in Sec.~\ref{sec:max_ent}.

When $T_2=0$, substantial entanglement is observed for any $T_1 \neq 0, 1$.
In this scenario, the collision between the initial electron and the first qubit in the steady state induces their entanglement the most when $\Omega_1=\pi f_{\text{col}}$ and $\theta_1=\pi/2$, see Appendix~\ref{A:rho1} and Eq.~\eqref{eq:Ce1}.
The strong collision (namely $T_2=0$) between the electron and the second qubit results in perfect spin exchange, described by $s_2 = -i ( \ket{\Up\up}\bra{\Up\up} + \ket{\Down\up}\bra{\Up\down} +\ket{\Up\down}\bra{\Down\up} +\ket{\Down\down}\bra{\Down\down})$, according to Eq.~\eqref{eq:sn}.
Hence, the entanglement originally present between the electron and the first qubit is transferred to entanglement between the two qubits.
The optimal conditions for the magnetic fields are $\Omega_1=\pi f_{\text{col}}$, $\theta_1=\pi/2$ for $\vec{B}_1$ while  $\vec{B}_2$ is irrelevant as it does not affect the entanglement generation in this scenario.

Conversely, when $T_1=0$, no entanglement is observed for any $T_2$.
Here, the electron and the first qubit are not entangled at all, as the strong collision between them leads the first qubit to align with the initial electron spin, as per Eq.~\eqref{eq:rho1-sc}.
Hence, the electron cannot mediate any entanglement between the qubits. 

In cases where $T_1\approx 1$ or $T_2\approx1$, the entanglement generation between the electron and the first (second, respectively) by each collision is small.
Hence, unless forming an entangled steady state as in the case $T_1 \approx T_2$ case, the entanglement vanishes.

When the electron-qubit interaction changes from ferromagnetic (the setup of Fig.~\ref{fig:ent_T1T2}) to antiferromagnetic type, the steady state entanglement is the same as Fig.~\ref{fig:ent_T1T2} when the magnetic field directions are changed as $\theta_n\rightarrow \pi-\theta_n$ and 
$\phi_n \rightarrow -\phi_n$ 
(while maintaining the field strengths $\Omega_n$ the same).
This symmetry operation ensures that the dynamics and resulting steady states of the antiferromagnetic case map directly onto those of the ferromagnetic case under time-reversal and spin-flip transformations.
Therefore, both cases exhibit identical entanglement properties.

\section{Discussion}
\label{sec:dis}
We show that coherent collisions involving itinerant electrons can effectively generate and stabilize entanglement between two remote  spin qubits.
Entanglement is generated during collisions where the electron exchanges spins coherently with either qubit.
By optimizing local magnetic fields, the collisions drive the qubits into an entangled steady state across a broad parameter space.
Particularly, when the collision strengths between the electron and each qubit are similar, the steady state entanglement achieves the maximum in the regime of weak collisions.
Interestingly, the optimal condition requires that the two fields should not only be perpendicular to the initial electron spin, but also differ themselves by an angle equal to the forward-scattering phase of the collisions. 
This optimal condition can be established in experiments by measuring the electron spin current at the output channel.

In contrast to recent studies, such as Li {\it et al.}\cite{li_effect_2023}, which require a temperature bias and system interactions absent in our model, we observe significant entanglement generation solely through coherent electron collisions. 
Another study~\cite{xiong_increasing_2008} consider the collisional model with depolarized electron spins, corresponding to our model with $p_U=0.5$, and concludes that the steady state entanglement is generated by the collisions. However, their reliance on von Neumann entropy of reduced density matrices for quantifying entanglement, which does not fully capture entanglement in mixed states, led to the conclusion which contrast with our emphasis on the crucial role of coherent electron spins for entanglement generation, as illustrated in Fig.~\ref{fig:ent_pU}.

M. Benito {\it et al.}~\cite{benito_dissipative_2016} have considered an electronic analog which approximates the dissipative entanglement generation mediated by photons~\cite{muschik_dissipatively_2011,krauter_entanglement_2011,stannigel_driven-dissipative_2012}. 
The master equation of the qubits is approximated to the one suggested in Ref.~\cite{stannigel_driven-dissipative_2012}, when the couplings between the qubits and the itinerant electron are symmetric and weak, and when the local magnetic fields are even weaker. 
Then sizable steady state entanglement is achieved, up to 0.7 of the entanglement of formation~\cite{wootters_entanglement_1998} which corresponds to $0.78$ of the concurrence.
The lack of full entanglement is due to the high-order effect of the coupling, which causes the deviation from the dynamics of the photonic setup.
On the other hand, our work does not prioritize to realize the master equation suggested in ~\cite{stannigel_driven-dissipative_2012}.
Instead, we have considered a generic collisional setup and searched optimal condition of the local drivings which maximizes the steady state entanglement without the approximations, thanks to the scattering formalism.
Our result shows that the full entanglement not only requires the symmetric and weak coupling, but also the strong transverse local fields (namely, $\Omega = \pi f_{\text{col}} \gg (1-T) f_{\text{col}}$) whose angle differ by the forward-scattering phase.

S. Sauer {\it et al.}~\cite{Sauer2013,Sauer2014} considered optimizing Hamiltonian for the two interacting qubits subject to given local decoherence channel to maximize the steady-state entanglement. In contrast, our work consider optimizing local Hamiltonians for two noninteracting remote qubits, subject to local decoherence and nonlocal dissipation mediated by the itinerant electrons. The mechanism responsible for the entanglement is the qubit interaction in the former and the nonlocal dissipation in the latter. Due to this difference, the maximal concurrence in our protocol exceeds the maximal stabilizable entanglement of $C=1/2$ found in Refs.~\cite{Sauer2013,Sauer2014}.

  We discuss the effect of fluctuation in the single-electron source to the performance of our protocol.  As analyzed in Ref.~\cite{Ryu2024}, single-electron sources of high energy types emit wave packets which are in identical form except picoseconds fluctuations in the emission timing.The standard deviation of the fluctuation, $\sigma_P$, is expected to be order of  picoseconds in the case of high-energy sources. Hence, the fluctuation of $f_{\text{col}}$ is estimated to be small, as $\delta f_{\text{col}}\sim (f_{\text{col}}^{-1}-\sigma_P )^{-1}-f_{\text{col}} \sim f_{\text{col}}^2 \sigma_P \sim 10^{-3} f_{\text{col}}$ for GHz sources.
Among the parameters to be optimized, only the strengths of local magnetic fields, $\Omega_1$ and $\Omega_2$, depend on $f_{\text{col}}$. One can roughly estimate the effect of fluctuation in $f_{\text{col}}$ by investigating the change of steady-state entanglement when $\Omega_1$ and $\Omega_2$ deviate from the optimal condition, Eq.~(\ref{eq:Omg}), by amount of $\pi \delta f_{\text{col}}$. We confirm that the change of the entanglement is negligible, $\delta C/C < 10^{-3} $ for $\delta f_{\text{col}}/f_{\text{col}}=10^{-3}$. The other model parameters, $T_1$ and $T_2$, do not depend on the emission timing. Hence, we expect that the small timimg noise will barely impact the performance of the protocol.

Implementing our proposed model is experimentally feasible. Decoherence rates of spin qubits are typically much slower than the collision rate $f_{\text{col}}$. For example, $0.1 \text{GHz} \le f_{\text{col}} \le 10$GHz in single-electron sources~\cite{feve_-demand_2007,giblin_towards_2012,dubois_minimal-excitation_2013}, while the relaxation ($T_1^{-1}$) and dephasing ($T_2^{-1}$) rates are in the orders of 0.1kHz and 1MHz, respectively, for GaAs quantum-dot spin qubits~\cite{bertrand_fast_2016,burkard_semiconductor_2023}. The condition for the coherent collision, namely that the energy uncertainty of the wave packet is much larger than the qubit level spacing, is feasible when using quantum-dot single-electron sources; 
the Zeeman splitting $E_{Z}= 26 \mu e\text{V}/\text{T}$ (obtained using $\mu_B=58 \mu e\text{V}/\text{T}$ and $g=0.44$ for GaAs~\cite{burkard_semiconductor_2023}) 
is much smaller than the energy uncertainty ($\sim 1$ meV~\cite{fletcher_continuous-variable_2019}) of the wave packet, in typical situations ($B\le 10$ T).
Furthermore, our protocol covers broad experimental setups, as it does not not rely on specific microscopic details.  The local drivings can be realized by any mechanisms, e.g., Rabi oscillations, and the electron-qubit interaction can be any type. Therefore, our protocol will help increase the connectivity of remote qubits (whose distance is limited by the spin coherence length ranging from $5\mu$m~\cite{flentje_coherent_2017} to 100$\mu$m~\cite{stotz_coherent_2005}) in generic solid-state devices, when the photonic interface~\cite{beukers_remote-entanglement_2024} is not available.

{\it Acknowledgement}---
We thank Patrick P. Potts, Gonzalo Manzano, and \.{I}nan\c{c} Adagideli for discussions.
We also thank Wilhelm and Else Heraeus-Foundation for organizing seminar "Entropy and the Second Law of Thermodynamics - The past, the present, and the future" which stimulated the initial idea.
R.L and S.R acknowledge support by the Spanish State Research Agency (MCIN/AEI/10.13039/501100011033) and FEDER (UE)
under Grant No.~PID2020-117347GB-I00 and Mar\'{i}a de Maeztu Project No.~CEX2021-001164-M and Grant No.~PDR2020/12 sponsored by the Comunitat Autonoma de
les Illes Balears through the Direcci\'{o} General de Pol\'{i}tica
Universitaria i Recerca with funds from the Tourist Stay
Tax Law ITS 2017-006.
A.M.B. acknowledges the Dutch Organization for Scientific Research (NWO) Grant No: OCENW.GROOT.2019.004.

\appendix

\section{Interaction between electrons and qubits}
\label{A:Heisenberg}

In this appendix, we consider the interaction between an electron spin and a spin qubit and show that how it determines the electron-qubit scattering amplitudes. Although here we consider a specific interaction model, its details such as spatial dependence or type of spin-spin coupling (ferromagnetic/antiferromagnetic), are irrelevant for the results in the main text.

The most dominant interaction between an electron and a spin qubit is given by the on-site Heisenberg exchange Hamiltonian~\cite{slichter1990Principles}:
\begin{equation}
\begin{aligned}
H_{int}(x) =\mp \lambda  \bm{\tau} \cdot\bm{\sigma}_n\delta(x-x_n),
\end{aligned}
\end{equation}
where $x$ denotes the electron position, $\bm{\tau}$ is the vector of Pauli spin matrices in electron spin subspace, $\bm{\sigma}_n$ is the vector of Pauli spin matrices for the spin qubit located at $x_n$ and $\lambda  \,(>0)$ is the effective interaction strength between electron spin and spin qubit. The sign is $-$ ($+$) when the interaction is ferromagnetic (antiferromagnetic).

Due to the assumptions that the electron travels with a constant velocity $v$ and that the initial electron's kinetic energy uncertainty is much larger than the qubits' level spacing (so that the spatial form of the electronic wavefunction remains unchanged by the exchange interaction), the time-dependent effective interaction between the electron and qubit becomes
\begin{equation}
    H(t) = \mp \lambda \bm{\tau} \cdot \bm{\sigma}_n |\psi(0;t)|^2,
\end{equation}
where $\psi(0;t)$ is the electronic wavefunction at the location of the qubit at time $t$.
Then the electron and qubit spins evolve in time according to the operator
\begin{align}
    U(t\rightarrow \infty) &= \exp \Big[ -\frac{i}{\hbar} \int_{-\infty}^\infty
    H(t) \, dt \Big] ,\\
    &= \exp \Big[ \mp \frac{i}{\hbar} \lambda \bm{\tau} \cdot \bm{\sigma}_n /v \Big] .
\end{align}
Using that $\bm{\tau}\cdot \bm{\sigma}_n$ is 1 for the triplets and -3 for the singlet, one obtains
\begin{align}
    \braket{\Down \down | U(t\rightarrow \infty)|\Down \down }
    &= e^{\mp i \lambda/(\hbar v)} , \label{eq:expPhi}\\
    \braket{\Down \up | U(t\rightarrow \infty)|\Down \up }
    &= \frac{1}{2} \Big[e^{\mp i \lambda/(\hbar v)} 
    +e^{\pm i 3\lambda/(\hbar v)} \Big] , \label{eq:t}\\
    \braket{\Up \down | U(t\rightarrow \infty)|\Down \up }
    &= \frac{1}{2} \Big[e^{\mp i \lambda/(\hbar v)} 
    -e^{\pm i 3\lambda/(\hbar v)} \Big] . \label{eq:r}
\end{align}
Eq.~\eqref{eq:expPhi}, Eq.~\eqref{eq:t}, and Eq.\eqref{eq:r} correspond to $e^{i\varphi_n}$ [Fig.~\ref{fig:setup}(b)], $t_n$ [Fig.~\ref{fig:setup}(c)] and $r_n$ [Fig.~\ref{fig:setup}(d)], respectively. 
Eqs.~\eqref{eq:expPhi}--\eqref{eq:r} give gauge-invariant relations among $e^{i\varphi_n}$, $t_n$, and $r_n$, 
\begin{align}
    e^{i\varphi_n}
    &= t_n+r_n , \\
    \arg{(r_n)} 
    &=  \arg{(t_n)} \mp \pi/2 .
\end{align}
    
\section{steady state of the first qubit}
\label{A:rho1}

Here we analyze the reduced density matrix of the first qubit in the steady state, which helps to find optimal conditions for the steady state entanglement between the two qubits.

Due to the causality, the reduced density matrix of the first qubit is determined by the collision with the incident electrons, irrelevantly to the subsequent collision with the second qubit. (This is analogous to the photonic setup of Ref.~\cite{stannigel_driven-dissipative_2012}.)
Hence, the reduced density matrix can be obtained from a situation simplified by detaching the second qubit, namely by setting $t_2=1$. 

We find that the reduced density matrix in the steady state $\rho^{(1)}_{\text{st}}$ is determined  as,
\begin{equation}
\begin{aligned}
&\braket{\up|\rho^{(1)}_{\text{st}}|\up} 
= \frac{1}{N}
\sin^2(\theta_1) \sin^2 (\Omega_1 f_{\text{col}}^{-1}/2) ,  \\    &\braket{\up|\rho^{(1)}_{\text{st}}|\down}= \frac{1}{N} \sin(\theta_1) \sin(\Omega_1 f_{\text{col}}^{-1}/2)   \\
&\, \times \Big[ i \Big(1- T_1 \pm i\sqrt{T_1(1-T_1)}\Big)\cos\Big(\frac{\Omega_1  f_{\text{col}}^{-1}}{2}\Big)   \\
&\, + \Big(-1- T_1 \pm i \sqrt{T_1(1-T_1)} \Big) \sin\Big(\frac{\Omega_1 f_{\text{col}}^{-1}}{2}\Big) \cos(\theta_1) \Big]  ,
\end{aligned}
\label{eq:rho1}
\end{equation}
$\braket{\down|\rho^{(1)}_{\text{st}}|\down} = 1-\braket{\up|\rho^{(1)}_{\text{st}}|\up}$, 
$\braket{\down|\rho^{(1)}_{\text{st}}|\up} = \braket{\up|\rho^{(1)}_{\text{st}}|\down}^*$,
and $N$ is the normalization factor,
\begin{equation}
\begin{aligned}
 & N=
  1+T_1 -\frac{T_1}{2} \big(3+\cos(2\theta_1)\big)\cos(\Omega_1 f_{\text{col}}^{-1})  \\
& \,  - T_1\sin^2(\theta_1) \pm 2\sqrt{T_1(1-T_1)}\cos(\theta_1) \sin(\Omega_1  f_{\text{col}}^{-1}) . 
\end{aligned}   
\end{equation}
We recall that $\theta_1$ is the polar angle of the magnetic field applied to the first qubit  (the azimuthal angle is assumed to be zero without loss of generality) and $\Omega_1 $ is the Larmor angular frequency.

In the weak collision limit, namely when $T_1 \rightarrow 1$, the reduced density matrix is approximated in the lowest order w.r.t. the collision strength as
\begin{equation}
\begin{aligned}
\rho_{\text{st}}^{(1)}
= &\frac{1}{3+\cos(2\theta_1)}
\begin{bmatrix}
    \sin^2(\theta_1) & -\sin(2 \theta_1) \\
     -\sin(2 \theta_1) & 1+3 \cos^2(\theta_1) 
\end{bmatrix} \\
&\quad +\mathcal{O}(\sqrt{1-T_1}).
\end{aligned}
\label{eq:rho1-wc}
\end{equation}
Eq.~\eqref{eq:rho1-wc} provide us the optimal direction of the magnetic field to form the maximal entanglement in the two-qubit steady state.
Namely, the optimal field should be perpendicular to the initial electron spin, $\theta_1=\pi/2$, to make the reduced density matrix the white-noise state, $\rho^{(1)}_{\text{st}} = \mathbb{1}/2$.
The higher-order corrections suggest the optimal field strength, $\Omega_1$.
For $\theta_1=\pi/2$ and general $T_1$, Eq.~\eqref{eq:rho1} is simplified as,
\begin{align}
    \braket{\up|\rho^{(1)}_{\text{st}}|\up} 
    =& \frac{\sin^2(\Omega_1 f_{\text{col}}^{-1}/2)}{1-T_1 \cos(\Omega_1 f_{\text{col}}^{-1}) } ,
     \\
    \braket{\up|\rho^{(1)}_{\text{st}}|\down}       
    =& \frac{\big\{ i(1-T_1) \mp \sqrt{T_1(1-T_1)} \big\} \sin(\Omega_1 f_{\text{col}}^{-1})  }{ 2[1-T_1 \cos(\Omega_1 f_{\text{col}}^{-1})]} .
\end{align}
Hence, $\Omega_1 f_{\text{col}}^{-1} =\pi $ is optimal, thereby causing maximal decoherence of the reduced density matrix, $\braket{\up| \rho^{(1)}_{\text{st}}|\down} =0$.

For arbitrary collision strength, 
we confirm that the optimal condition remains the same as the above prediction, namely $\theta=\pi/2$ and $\Omega f_{\text{col}}^{-1}=\pi$. 
Via numerical calculation using Eq.~\eqref{eq:rho1}, we observe that the purity of the reduced density matrix, $\text{Tr}\,(\rho^{(1)}_{\text{st}})^2$, becomes minimal at $\theta=\pi/2$ and $\Omega f_{\text{col}}^{-1}=\pi$. 
This is plausible because the corresponding Larmor precession, namely the spin flip along the z-direction, prohibits the most the collisions from polarizing the qubit to $\ket{\down}$.
In the optimal condition, the concurrence of the bipartite entanglement of electron and the first qubit, whose joint state is described by $\rho^{(e,1)}$,  becomes
\begin{equation}
    C(\rho^{(e,1)})
    = \frac{4(1-T)T}{(1+T)^2} .
    \label{eq:Ce1}
\end{equation}

In the strong collision limit, namely when $T_1 \rightarrow 0$, the entanglement $C(\rho^{(e,1)})$ vanishes. 
In such limit, the reduced density matrix approaches to a pure state (which is as expected from the vanishing entanglement),
\begin{equation}
    \rho^{(1)}_{\text{st}}
    =  e^{\frac{i}{2}\Omega f^{-1}_{\text{col}} \bm{\sigma}_1 \cdot \hat{B}_1} \ket{\down}\bra{\down} 
     e^{-\frac{i}{2}\Omega f^{-1}_{\text{col}} \bm{\sigma}_1 \cdot \hat{B}_1} .
     \label{eq:rho1-sc}
\end{equation}
This is because the strong collision makes the qubit to polarize along the initial electron spin, $\ket{\Down}$, and the following Larmor precession rotates the qubit according to the unitary operation, $\exp(i \Omega  f^{-1}_{\text{col}} \bm{\sigma}_1\cdot \hat{B}_1 /2)$.

\section{Singular value decomposition of collision}
\label{A:SVD}

Here we present the singular value decomposition of the electron-qubits scattering operator $S$ and the derivation of Eqs.~\eqref{eq:tt} and \eqref{eq:rt}.
When the strength of the collision is symmetric for the two qubits, namely when $T_1=T_2\equiv T$ (hence $\varphi_1=\varphi_2\equiv \varphi$), we find that
\begin{equation}
\begin{aligned}
    \braket{\Down | S | \Down}
    =& e^{2i \varphi} \ket{\down\down}\bra{\down \down}+ T \ket{\up\up}\bra{\up\up} \\
    &+T e^{i \varphi} \ket{\Psi(\vartheta_T, -\varphi)} \bra{\Psi(\pi-\vartheta_T, -\varphi )} \\
    &+e^{i\varphi} \ket{\Psi(\pi-\vartheta_T, \pi-\varphi)}\bra{\Psi(\vartheta_T, \pi-\varphi)} ,
\end{aligned}
\label{eq:Sdd-svd}
\end{equation}
\begin{equation}
\begin{aligned}
    \braket{\Up | S | \Down}
    =& \mp i\sqrt{1-T^2} \ket{\down\down}
    \bra{\Psi(\pi-\vartheta_T, -\varphi)}  \\
    & \mp i\sqrt{1-T^2} \ket{\Psi(\pi-\vartheta_T , \varphi) } \bra{\up\up} .
\end{aligned}
\label{eq:Sud-svd}
\end{equation}
We recall that $\vartheta_T \equiv 2\, \text{atan} (\sqrt{T})$ and 
$
    \ket{\Psi(\vartheta, \varphi)  }  \equiv
    \cos(\vartheta/2) \ket{\up\down}
    +\sin(\vartheta/2) e^{i \varphi} \ket{\down \up} .
$
The equations~\eqref{eq:tt} and \eqref{eq:rt} are derived when using Eqs.~\eqref{eq:Sdd-svd}, \eqref{eq:Sud-svd}, and the following relations,
\begin{align}
& 
e^{-i\frac{H}{\hbar f_{\text{col}}}} 
\ket{\down\down}
=-e^{-i\phi}\ket{\up\up} , \\
& e^{-i\frac{H}{\hbar f_{\text{col}}}} \ket{\up\up}
=-e^{i\phi}\ket{\down\down} , \\
& e^{-i\frac{H}{\hbar f_{\text{col}}}}\ket{\Psi(\vartheta,\varphi)}
=-e^{i(\varphi+\phi)} \ket{\Psi(\pi -\vartheta, -\varphi-2\phi)} ,
\end{align}
where $H$ is determined by Eqs.~\eqref{eq:H} and \eqref{eq:B1}--\eqref{eq:Omg}.

\begin{figure}[h]
  \centering
  \includegraphics[width=\columnwidth]{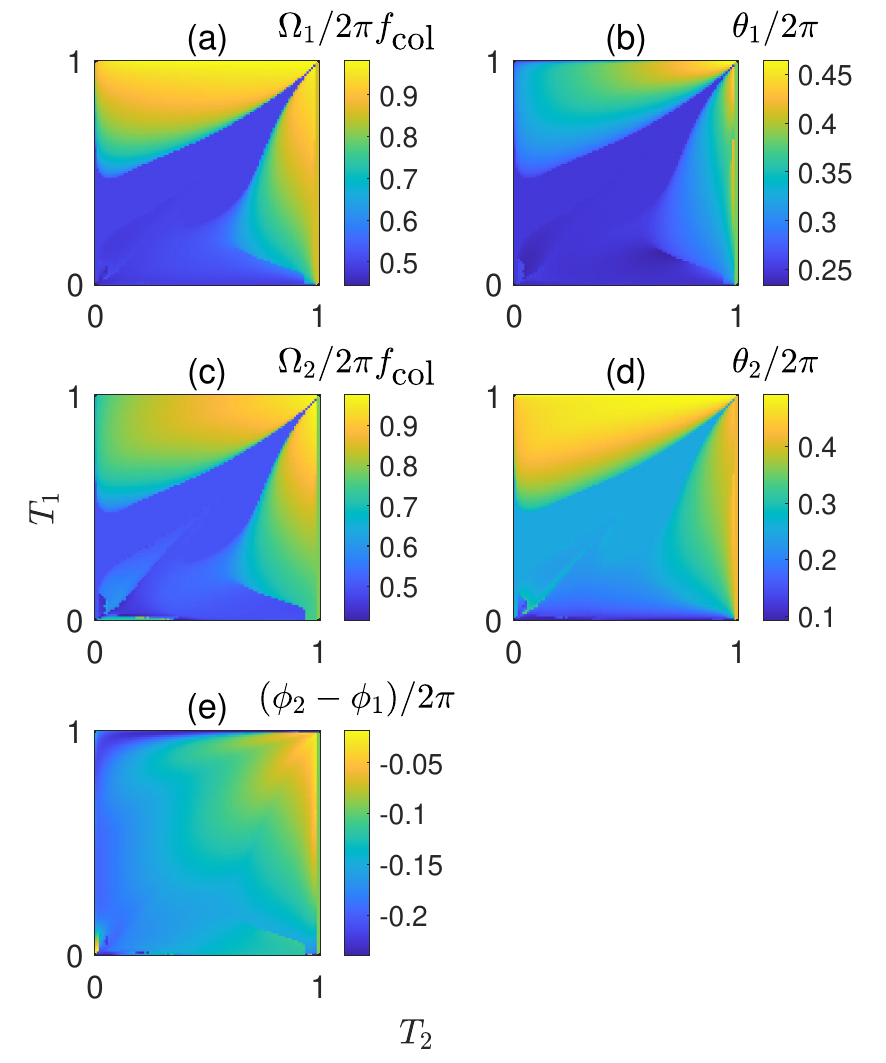} 
  \caption{Parameters of the magnetic fields for Fig.~\eqref{fig:ent_T1T2}. }
  \label{fig:pam_ent_T1T2}  
\end{figure}

\section{Parameters of magnetic fields for Fig.~\eqref{fig:ent_T1T2}}

Figure \ref{fig:pam_ent_T1T2} shows the parameters of the magnetic fields which was used to obtain the steady state entanglement in Fig.~\ref{fig:ent_T1T2}.
Theses are the results of the numerical optimization maximizing the steady state entanglement. 
Small fluctuations around $T_1,T_2=0$ are due to the hopping among multiple solutions giving the same entanglement. 


%

\end{document}